\documentclass[aps,prb,twocolumn,showpacs,showkeys]{revtex4}

\usepackage{graphicx,color,subfig,amsmath,bm,amssymb}
\usepackage[USenglish]{babel}

\begin{document}

\title{Calculated Cleavage Behavior and Surface States of LaOFeAs}

\author{Helmut Eschrig, Alexander Lankau, Klaus Koepernik}

\email{k.koepernik@ifw-dresden.de}
\homepage{http://www.ifw-dresden.de/~/magru}

\affiliation{IFW Dresden, PO Box 270116, D-01171 Dresden, Germany}

\begin{abstract}
The layered structure of the iron based superconductors gives rise to a
more or less pronounced two-dimensionality of their electronic
structure, most pronounced in LaOFeAs. A consequence are distinct
surface states to be expected to influence any surface sensitive
experimental probe. In this work a detailed density functional analysis
of the cleavage behavior and the surface electronic structure of LaOFeAs
is presented. The surface states are obtained to form two-dimensional
bands with their own Fermi surfaces markedly different from the bulk
electronic structure. 
\end{abstract}

\pacs{68.35.bd, 74.20.Pq, 74.25.Jb, 74.70.Db}
\keywords{surface relaxation, surface band structure}

\maketitle

\section{Introduction}

About two years after the discovery of the high-T$_c$ superconductor
La(O,F)FeAs\cite{kamihara} its normal state electronic structure remains
badly understood. Although there is growing opinion that it is not a
highly correlated case of Mott-Hubbard type\cite{anisimov}, its magnetic
properties for instance are not at all well described by the local
density approximations (LDA) of density functional theory (DFT) or
gradient corrections to it (GGA). In contrast to the cuprates where it
was nearly immediately clear that the electronic structure of the
cuprate plane is dominated by a $3d^9$ shell with a correlation
localized hole in a single band, here a complex multi-band case is
obviously operative.\cite{singh} Like in the cuprates the electronic
structure and most properties are highly anisotropic, though to varying
extent.\cite{tb}

Many experimental probes of the electronic structure, photoemission in
particular, are surface sensitive. To assess these experiments it is
crucial to know something on the surface structure and possible surface
states of the material. So far, as a rule those experiments are
interpreted in terms of the bulk electronic structure. Even if bulk
properties show up in the experiment, the possible presence of surface
states remains an important issue to be taken into account.

In this paper the formation of surfaces in cleaving a single crystal and
their properties are analyzed by means of DFT calculations for the most
two-dimensional case, undoped LaOFeAs. A pronounced surface electronic
structure is found for both forming surfaces, As and La terminated,
which dramatically differs from the bulk electronic structure. All
calculations are done within the non-magnetic GGA model of LaOFeAs which
fairly correctly describes the crystal structure of the
material. Although the relevance of this model for electronic many-body
behavior may be debated, the results on the surface relaxation and on
the character of surface states must be considered relevant.

The computational approach is described in the following section, while
the results are presented and discussed in Section III. A short summary
completes the paper.

\section{Computational Approach}

All calculations were done by using the full-potential local-orbital
(FPLO) code version FPLO9.00.\cite{FPLO} Relativistic effects were
incorporated on a scalar relativistic level. The $\bm k$-mesh input
parameters were set to 6,6,4 (number of $\bm k$ intervals along the
$a,\, b,$ and $c$ axes of the Brillouin zone) for bulk calculations and
to 6,6,2 for slab supercell calculations. Some calculations were
repeated with parameters 12,12,4 and 12,12,2, resp., with essentially no
changes in the results. The density convergence parameter was always put
to $10^{-6}$.

Since the surface structures were relaxed before calculating the surface
Kohn-Sham band structure, relaxed bulk reference structures had to be
used for comparison. The obtained structure parameters together with
experimental values are shown in Table~\ref{tab:1.1}. Since as usually
the GGA\cite{GGA} results are closer to experiment than the
LDA\cite{LDA} results, the GGA functional was used in all further
calculations. In this text tetragonal structures are considered only,
therefore the low temperature experimental structure data from Ref.\
\onlinecite{struc} were tetragonally averaged in Table~\ref{tab:1.1}. As
usual, $z_\text{O}$ was put to zero and $z_{\text{Fe}}$ to 1/2. We are
interested in the (001) surfaces of bulk crystals, so the $a$ lattice
constant is kept fixed at the GGA bulk value $a = 4.041$ \AA\ in all
what follows. In this text, the tetragonal axis of the structure is
referred to as $c$-axis and its direction as $z$-direction.

\begin{table}[h]
  \centering
  \begin{tabular}{|l|c|c|c|c|}
    \hline    &$\quad a[$\AA$] \quad$&$\quad c[$\AA$] \quad$&
               $\quad z_{\text{As}} \quad$&$\quad z_{\text{La}} \quad$
               \\
    \hline
    experim.  &  4.027  &  8.718  &  0.6513 &  0.1417 \\
    GGA       &  4.041  &  8.574  &  0.6403 &  0.1462 \\
    LDA       &  3.957  &  8.339  &  0.6410 &  0.1488 \\\hline
  \end{tabular}
  \caption{Structure parameters of bulk LaOFeAs}
  \label{tab:1.1}
\end{table}

\begin{figure}[h]
  \centering
  \includegraphics[angle=-90,scale=0.35,clip=]{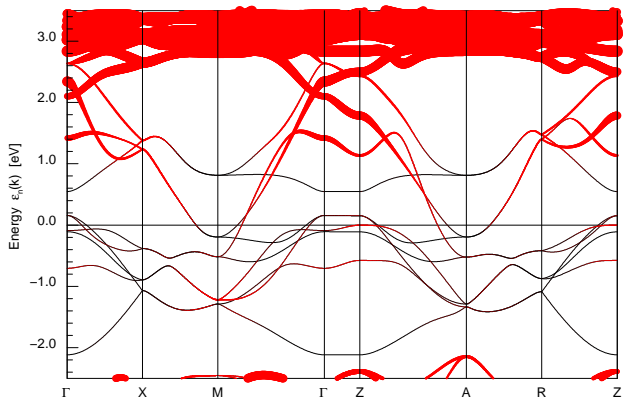}
  \caption{(color online) The Kohn-Sham GGA band structure of bulk
    LaOFeAs. The `fat bands' in red weigh the O-$2p$ and La-$4f,5d,6s$
    orbital contribution to the band state.}
  \label{fig:bulkbands}
\end{figure}

\begin{figure}[h]
  \centering
  \includegraphics[scale=0.2,clip=]{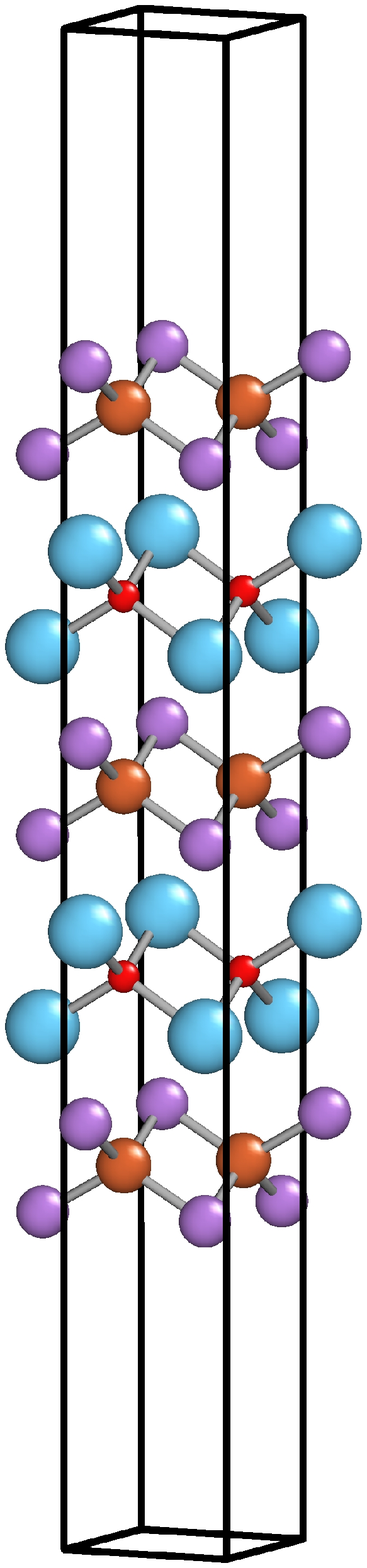}
  \includegraphics[scale=0.2,clip=]{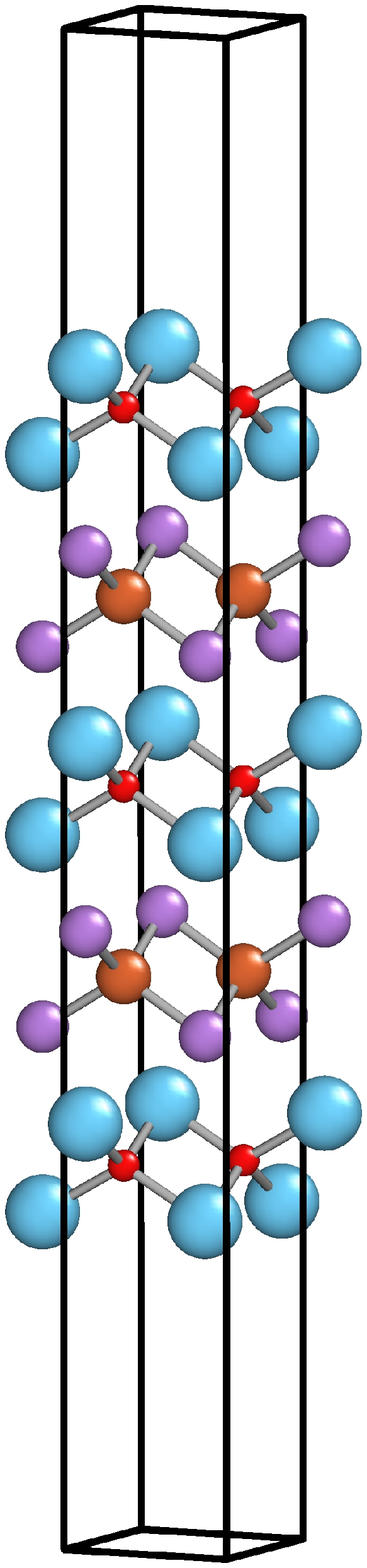}
  \caption{(color online) Slab geometry: unit cells of
    (LaO)$_4$(FeAs)$_6$ (left) and (LaO)$_6$(FeAs)$_4$ (right) slabs;
    blue: La, red: O, orange: Fe and violet: As}
  \label{fig:f1}
\end{figure}

The Kohn-Sham band structure of the GGA relaxed bulk crystal is shown in
Fig.~\ref{fig:bulkbands}. The `fat bands' (online in red) on this figure indicate
the O-$2p$ and La-$4f,5d,6s$ orbital character of the bands. The
thickness of those fat bands is proportional to the sum of squares of
coefficients for these basis states in the expansion of the Kohn-Sham
wavefunction into the FPLO orbital basis. The Fermi level is put to zero
in all presented band structures in this work. As is seen, the
conduction states at Fermi level and in the occupied region of the
conduction bands have very small wavefunction amplitudes in the LaO
layers (cf.\ Fig.~\ref{fig:f1} for the layered structure). As has
already been stated many times, they have predominantly Fe-$3d$
character and decay exponentially into the LaO layers. Experimentally it
is found\cite{mueller} that superconducting F-doped LaOFeAs shows an
intrinsic Josephson effect like bismuth cuprate\cite{Bi} which is only
possible if the spacer layers are not metallic and form tunnel barriers
instead. The band structure analysis of Fig.~\ref{fig:bulkbands}
indicates a conduction gap of several eV of the LaO layer. This is in
contrast to Ba(FeAs)$_2$ where the Ba layers are obtained at least close
to be metallic through Ba-$5d$ orbitals.\cite{tb} Also seen in the
figure by comparison of its left ($k_z = 0$) with its right part ($k_z
= \pi$) is the pronounced two-dimensional (2D) character of the
conduction states.

We calculated also the bands for the experimental structure parameters.
The deformation potentials with respect to change of the lattice
constants and of the Wyckoff parameters for the Fe-$3d$ bands crossing
the Fermi level are typically 2 eV/\AA. Both band structures deviate
from each other less than 0.2 eV in the window $\pm 1$ eV around Fermi
level, which is irrelevant for our considerations. (The two nearly
parallel dispersive bands along $\Gamma$-Z below Fermi level have
Fe-$3d_{z^2}$ character, which will appear even somewhat lower in
energy if the experimental Wyckoff parameter is used instead of the
GGA relaxed value (cf.  Table~\ref{tab:1.1}).) The unoccupied
dispersionless La-$4f$ bands are seen about 3 eV above Fermi level.

First we want to learn something on the cleavage behavior and the layer
bonding. To this goal, a periodic stacking of layers\\[1ex]
\centerline{As/Fe/As/La/O/La/As/Fe/As/La/O/La}\\[1ex]
is considered (in $c$-direction doubled unit cell). The symmetry is
lowered in this case to the orthorhombic space group Pmm2 so that all
atom layers have free $z$ Wyckoff parameters. (In the space group P4/nmm
of the bulk crystal the $z$ Wyckoff parameters of O and Fe layers are
fixed by symmetry to the values 0 and 1/2, respectively.) Now, for a
sequence of increasing $c$ lattice constants the $z$ Wyckoff parameters
of all atom layers were relaxed. As will be seen from the results
presented in the next section, the crystal is expected to cleave between
As and La layers only. Hence, clean coherent As or La surfaces are
expected after cleavage. In a real crystal, due to defects of course
terraces will likely be obtained consisting of coherent As and La
terminated areas, respectively.

Next, two types of symmetrically terminated periodically repeated slabs
with the full space group P4/nmm are considered with atom layer
stacking\\[1ex] 
\centerline{As/Fe/As/La/O/La/As/Fe/As/La/O/La/As/Fe/As}\\[1ex]
and\\[1ex]
\centerline{La/O/La/As/Fe/As/La/O/La/As/Fe/As/La/O/La}\\[1ex]
(Fig.~\ref{fig:f1}) with sufficiently large space between terminating
surfaces so that subsequent slabs do not interact any more.  For stacks
of symmetric slabs this is easily obtained if there is sufficient space
for the electronic states not to overlap across the spacing: Due to
symmetry there is no electric field in the free space between the slabs.
Since there are two atoms per Fe and O layer in the unit cell and one of
the other atoms per layer, the unit cells of these stacks are
(LaO)$_4$(FeAs)$_6$ and (LaO)$_6$(FeAs)$_4$, respectively.

To check that the results are representative for surfaces of bulk
crystals, both the spacing between the slabs and the slab thickness were
varied. In addition to the above, (LaO)$_8$(FeAs)$_{10}$ and
(LaO)$_{10}$(FeAs)$_8$ slabs are considered. Besides the results of
structure relaxation, stability of layer charges is considered as a
check for convergence of the considered slabs towards open bulk crystal
surfaces. Given the total charge density of a crystal there is ambiguity
to assign charges to the atom sites, and different codes differ in doing
so (if they make such an assignment at all). However, with such a charge
partitioning procedure fixed, the stability of the assigned charges with
respect to slab and spacer thickness can be taken as a criterion for
size convergence. Here, the site charges recorded in the FPLO protocol
file are used. As a result of these checks, while the slab
(LaO)$_4$(FeAs)$_6$ turned out to be representative, the thicker slab
(LaO)$_{10}$(FeAs)$_8$ had to be considered for the La terminated case.

Finally, for the geometrically relaxed slabs Kohn-Sham band structures
were calculated and projected onto various basis orbitals of the
`chemical basis' used in the FPLO code. This allows to distinguish bulk
from surface bands and to estimate the spatial extension of the
corresponding states. We mention here that FPLO works with a well
designed small basis consisting at most of one `chemical' basis state
(site orbital) and one polarization state per atomic orbital. The
chemical relevance of the `chemical' part of the basis is emphasized by
the fact that the polarization states are typically occupied by less
than 0.01 electron in a converged calculation, although relaxed
structures and total energies compete well in accuracy with any
available full potential density functional code like for instance
FLAPW.

Geometry relaxation was done by direct calculation of the forces on atom
sites and minimizing them with the corresponding tool of FPLO9.00.
Convergence was accepted when all forces were smaller than $10^{-3}$
eV/\AA. Only for the thickest slabs this was hard to achieve, and the
relaxation was stopped when the atom positions did not change any more
within 0.01 \AA\ accuracy over several force steps (in an oscillating
behavior) which keeps the band energies close to Fermi level within
about 20 meV of accuracy.

\section{Results and Discussion}

\subsection{Cleavage behavior}

Cleavage is ideally considered as expanding a bulk crystal in
$z$-direction until it disintegrates into parts. On the computer, a bulk
crystal with a unit cell doubled in $z$-direction and with the in-plane
lattice constant fixed to its equilibrium bulk value $a = 4.041$ \AA,
but with reduced space group to the orthorhombic group Pmm2 and with an
increasing sequence of $c$ lattice constants is treated. Pmm2 (still
with $b=a$) is the highest symmetry of our situation which leaves all
$z$ Wyckoff parameters of all atom layers undetermined. For each value
of the $c$ lattice constant these Wyckoff parameters were relaxed in
GGA.

\begin{figure}[h]
  \centering
  \includegraphics[scale=0.6,clip=]{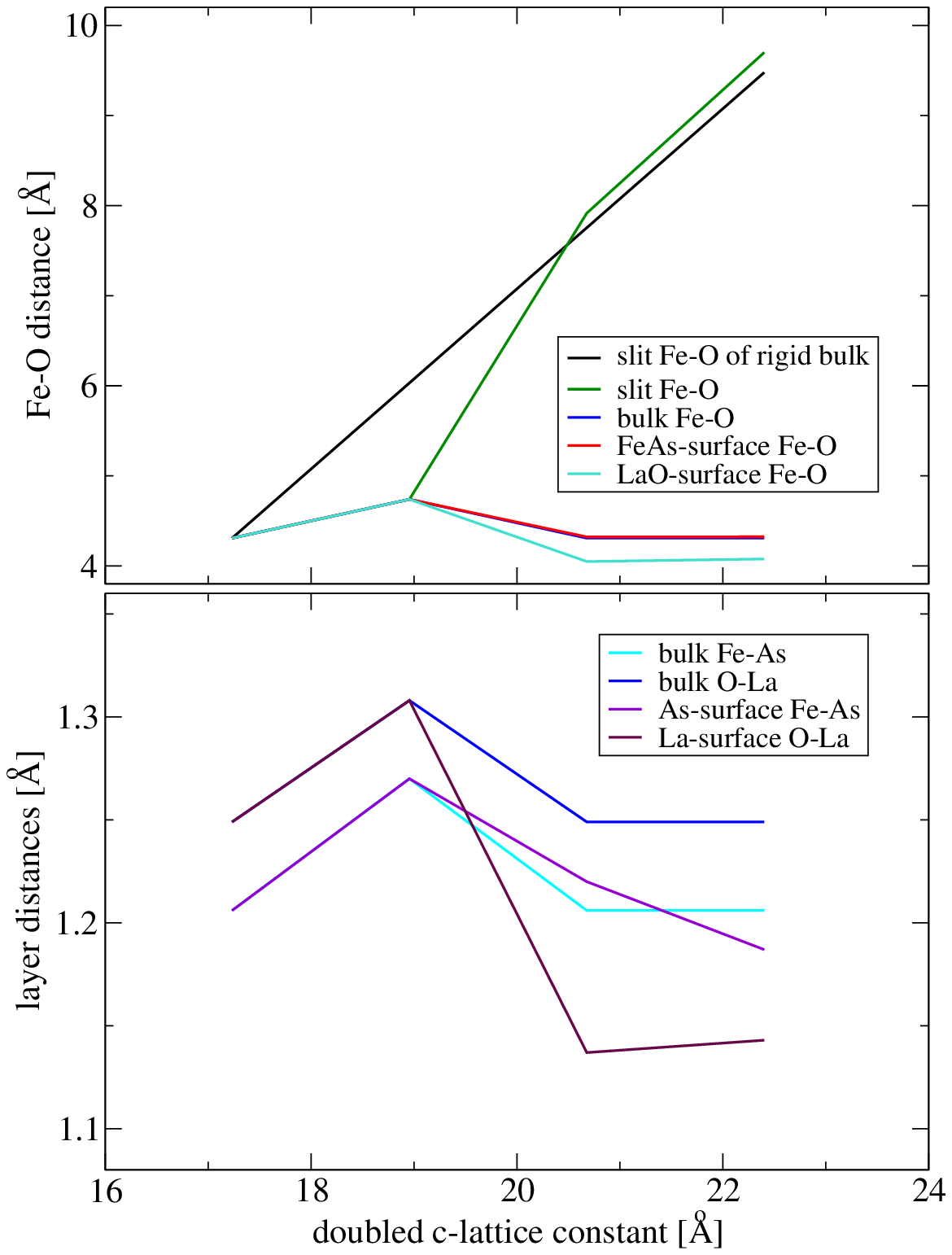}
  \caption{(color online) The atom layer distances as functions of the
    stretched $c$ axis lattice constant. In the upper panel the Fe-O
    layer distances are shown, the two upper curves across the slit and
    the two lower curves between the surface and subsurface triple layer
    on both sides of the slit (FeAs and LaO terminated, respectively. On
    the lower panel the behavior of layer distances within the triple
    layers La/O/La and As/Fe/As, respectively, are shown, again for the
    bulk and for the surface triple layers. Observe the different scales
    of the ordinates. For more explanation see the text.}
  \label{fig:f2}
\end{figure}

The crystal structure of LaOFeAs can be imagined as a stacking of layers
of tetrahedra with their corners occupied with La and As
sites, respectively, and which are centered by O and Fe ions (triple atom
layers, Fig.~\ref{fig:f1}). If the bulk structure of the crystal were
rigid and the crystal would split between every fourth La-As layer
pair, the slit width as measured between the Fe and O layers forming the
centers of the adjacent tetrahedra layers would be that of the black
straight line in the upper panel of Fig.~\ref{fig:f2}. For $z$-strain
not exceeding about 10 p.c. (leftmost line segments in Fig.\
\ref{fig:f2}) the tetrahedra behave indeed as nearly rigid, but every
La-As distance is homogeneously strained by one fourth of the black
line. Observe that the ordinate scale is stretched by about a factor of
20 in the lower panel of Fig.~\ref{fig:f2} compared to the upper, so
the changes in the La-O and Fe-As distances within the triple layers are
really negligible compared to the change of the distance between the
triple layers (smaller roughly by a factor of seven). Hence, for not to
large strain the crystal behaves as a system of nearly rigid triple atom
layers La/O/La and As/Fe/As which are elastically bonded together.

\begin{table*}[t]
  \centering{\footnotesize
  \begin{tabular}{|c|c|c|c|c|c|}
    \hline
            &        &        &        &        &         \\
            &\hspace*{2.2em}bulk\hspace*{2.2em}&
                     (LaO)$_4$(FeAs)$_6$&(LaO)$_8$(FeAs)$_{10}$&
                     (LaO)$_6$(FeAs)$_4$&(LaO)$_{10}$(FeAs)$_8$  \\
            &        &        &        &        &         \\
    \hline
            &        &        &        &        &         \\[-2ex]
    $\cdots$&$\cdots$&$-----$ &$-----$ &        &         \\
    As/Fe   & 1.203  & 1.146  & 1.150  &        &         \\
    Fe/As   & 1.203  & 1.161  & 1.168  &        &         \\
    As/La   & 1.830  & 1.815  & 1.830  &$-----$ &$-----$  \\
    La/O    & 1.254  & 1.249  & 1.255  & 0.940  & 0.948   \\
    O/La    & 1.254  & 1.244  & 1.248  & 1.536  & 1.529   \\
    La/As   & 1.830  & 1.799  & 1.832  & 1.420  & 1.436   \\
    As/Fe   & 1.203  & 1.193  & 1.202  & 1.198  & 1.206   \\
    Fe/As   & 1.203  &$\cdots$& 1.201  & 1.179  & 1.168   \\
    As/La   & 1.830  &        & 1.831  & 1.803  & 1.871   \\
    La/O    & 1.254  &        & 1.253  & 1.243  & 1.208   \\
    O/La    & 1.254  &        & 1.254  &$\cdots$& 1.263   \\
    La/As   & 1.830  &        & 1.831  &        & 1.729   \\
    As/Fe   & 1.203  &        & 1.202  &        & 1.172   \\
    Fe/As   & 1.203  &        &$\cdots$&        & 1.179   \\
    As/La   & 1.830  &        &        &        & 1.696   \\
    La/O    & 1.254  &        &        &        & 1.222   \\
    $\cdots$&$\cdots$&        &        &        &$\cdots$ \\[1ex]
    \hline
  \end{tabular}}
  \caption{GGA relaxed interlayer distances in \AA\ for the bulk crystal
    and for various slabs. In view of the mirror symmetry only half of
    each slab is presented: its center is at the lower end and the
    surface is indicated by a dashed line.} 
  \label{tab:dist}
\end{table*}

\begin{table*}[tbh]
  \centering{\footnotesize
  \begin{tabular}{|c|l|l|l|l|l|l|}
    \hline
          &        &        &        &        &       &  \\
    layer &\hspace*{3.2em}bulk\hspace*{3.2em}&
    (LaO)$_4$(FeAs)$_4$:2.6c&(LaO)$_4$(FeAs)$_6$:4c&
    (LaO)$_4$(FeAs)$_6$:6c&(LaO)$_8$(FeAs)$_{10}$:6c&
    (LaO)$_{10}$(FeAs)$_8$:6c \\
          &        &        &        &        &        &  \\
    \hline
          &        &        &        &        &        &  \\[-2ex]
          &$\;\cdots$&$\;------$&$\;------$&$\;------$&$\;------$&  \\
    As    &$\;-0.738$&$\;-0.358$&$\;-0.179$&$\;-0.179$&$\;-0.177$&  \\
    Fe$_2$&$\;+0.287\;(-1.189)$&$\;+0.392\;(-0.731)$&$\;+0.299\;(-0.611)$
          &$\;+0.297\;(-0.608)$&$\;+0.307\;(-0.609)$& \\
    As    &$\;-0.738$&$\;-0.765$&$\;-0.731$&$\;-0.726$&$\;-0.735$&$\;------$ \\[1ex]
    La    &$\;+1.954$&$\;+1.959$&$\;+1.955$&$\;+1.955$&$\;+1.955$
          &$\;+1.467$ \\
    O$_2$ &$\;-2.719\;(+1.189)$&$\;-2.701\;(+1.215)$&$\;-2.706\;(+1.199)$
          &$\;-2.707\;(+1.198)$&$\;-2.706\;(+1.199)$
          &$\;-2.436\;(+0.786)$ \\
    La    &$\;+1.954$&$\;+1.947$&$\;+1.950$&$\;+1.950$&$\;+1.950$
          &$\;+1.755$ \\[1ex]
    As    &$\;-0.738$&$\;-0.726$&$\;-0.722$&$\;-0.719$&$\;-0.734$
          &$\;-0.840$ \\
    Fe$_2$&$\;+0.287\;(-1.189)$&$\;+0.272\;(-1.229)$&$\;+0.267\;(-1.177)$
          &$\;+0.258\;(-1.180)$&$\;+0.291\;(-1.182)$
          &$\;+0.233\;(-1.303)$ \\
    As    &$\;-0.738$&$\;-0.777$&$\;-0.722$&$\;-0.719$&$\;-0.739$
          &$\;-0.696$ \\[1ex]
    La    &$\;+1.954$&$\;+1.849$&$\;+1.950$&$\;+1.950$&$\;+1.954$
          &$\;+1.954$ \\
    O$_2$ &$\;-2.719\;(+1.189)$&$\;-2.445\;(+0.767)$&$\;-2.706\;(+1.199)$
          &$\;-2.707\;(+1.198)$&$\;-2.719\;(+1.187)$
          &$\;-2.727\;(+1.172)$ \\
    La    &$\;+1.954$&$\;+1.363$&$\;+1.955$&$\;+1.955$&$\;+1.954$
          &$\;+1.945$ \\[1ex]
    As    &$\;-0.738$&$\;------$&$\;-0.731$&$\;-0.726$&$\;-0.739$
          &$\;-0.738$ \\
    Fe$_2$&$\;+0.287\;(-1.189)$&                    &$\;+0.299\;(-0.611)$
          &$\;+0.297\;(-0.608)$&$\;+0.288\;(-1.190)$
          &$\;+0.192\;(-1.252)$ \\
    As    &$\;-0.738$&          &$\;-0.179$&$\;-0.179$&$\;-0.739$
          &$\;-0.707$ \\[1ex]
    La    &$\;+1.954$&          &$\;------$&$\;------$&$\;\cdots$
          &$\;+1.956$ \\
    O$_2$ &$\;-2.719\;(+1.189)$&          &          &          &          
          &$\;-2.718\;(+1.194)$ \\
    La    &$\;+1.954$&          &          &          &          
          &$\;+1.956$ \\
          &$\;\cdots$&          &          &          &          
          &$\;\cdots$ \\[1ex]
    \hline
  \end{tabular}}
\caption{FPLO GGA cross layer charges in (positive) elementary
  charges per atom of bulk and cleaved (second column) LaOFeAs and of
  various slabs of it. The number after the colon in the headings gives 
  the slit thickness in multiples of bulk $c$ lattice constants. Only
  the upper half of the last symmetric slab is presented. In parentheses
  the charges of the covalently bound triple layers are given.} 
  \label{tab:charges}
\end{table*}

For a sufficiently small $c$-strain of the crystal the original
periodicity is preserved. However, as said the lattice essentially
expands only between La and As layers while the La-O and As-Fe layer
distances hardly change. There are four neighboring La-As layer pairs in
the doubled unit cell. For larger strain the original periodicity breaks
up; some of the La-As layer distances snap back essentially to the bulk
value while slits open at others. By the symmetry restriction of our
approach (doubled periodicity in $c$-direction), adopted to keep the
computer time within reasonable limits, every fourth La-As spacing must
open a slit. These results are shown on Fig.~\ref{fig:f2}. 

Somewhere between 10 and 20 p.c. $z$-strain (we only calculated in 10
p.c. steps) this transition happens and the crystal cleaves.
Because of our periodicity constraint it must simultaneously cleave
between every fourth La-As layer pair. This behavior is seen in the
middle straight line segments of Fig.~\ref{fig:f2}. The layer distances
where the crystal does not cleave snap back to approximately their
equilibrium values and the distance at the cleavage slit approaches the
black line.  In truth this change is to be expected to happen abruptly
somewhere between 10 and 20 p.c. strain, however, the actual value is not
so interesting for our consideration. The layer distances do not exactly
go back to their bulk values because the surface triple layers on both
sides of the slit (one La/O/La and one As/Fe/As triple layer) relax a
bit. As is seen the relaxation is much larger on the La side compared to
the As side.

The clear message of these results is that LaOFeAs can only cleave
between La and As while the La and As tetrahedra are very stiff and
stable. Also, in the elastic deformation regime the crystal is not only
elastically extremely anisotropic, its unit cells also do not deform
homogeneously under uniaxial stress as the bonds between triple layers of
tetrahedra are almost one order of magnitude weaker than the bonds within
the tetrahedra.

\subsection{Surface relaxation}

The approach of the previous section is correct as long as the width of
the opening slit is still small enough so that both sides see each
other: Because the produced slabs do not have a mirror plane, they
produce a dipole charge density and hence electric fields across the
slit. This does not describe the situation of a single surface of a
crystal without its counterpart. As long as the sample is not charged
this surface does not produce a far reaching electric field. This
situation is better modeled by a symmetric slab of sufficient thickness
to represent a half-crystal. Now, however, two different slabs with
different possible surfaces have to be considered: As terminated and La
terminated surfaces.

In order to model an As terminated surface, periodically repeated slabs
(LaO)$_4$(FeAs)$_6$ and (LaO)$_8$(FeAs)$_{10}$ are considered with a
separation of the slabs between about 13 \AA\ and 30 \AA. These
structures have again the space group P4/nmm like the bulk crystal, with
a diagonal glide mirror plane in the middle. Therefore, these slabs do
not produce gross dipole densities of the slab (of course, as any
surface they have surface dipole densities, but here opposite to each
other on both surfaces), and no electric fields are produced in the free
space separating the slabs. They would only interact via wave function
overlap across the slit. These slab structures were now relaxed using
the forces on atom sites computed with the FPLO code. The obtained
surface relaxation does practically not change, if the spacing between
(LaO)$_4$(FeAs)$_6$ slabs is increased from 13 \AA\ to 30 \AA, and it
stays within 0.02 \AA, about twice the numerical accuracy of the
calculation, between the (LaO)$_4$(FeAs)$_6$ and (LaO)$_8$(FeAs)$_{10}$
slabs.

As already in the cleavage process, the La terminated surface is more
relaxing than the As terminated surface. Moreover, it turns out that the
(LaO)$_6$(FeAs)$_4$ slab is not thick enough in this case. The surface
triple layer dopes more charge into the neighboring FeAs triple layer
raising there the Fermi level against its bulk value. Besides the
surface potential step, there are two more essential influences on the
electronic structure of a slab: structure relaxation and a chemical
potential shift due to surface states. The latter is not operative on
the surface of a bulk crystal where the chemical potential (Fermi level)
is fixed to its bulk value. For this to be approximately true the
(LaO)$_6$(FeAs)$_4$ slab turns out to be not thick enough. Therefore, we
concentrate on a (LaO)$_{10}$(FeAs)$_8$ slab.

Table~\ref{tab:dist} shows the relaxed interlayer distances in
comparison with their bulk values. For the As terminated slabs,
essentially only the topmost triple layer relaxes by reducing the As-Fe
layer distances by about 0.05 \AA\ while all other distances inside
triple layers stay at their bulk values within 0.01 \AA\
accuracy. (Again, the bond between LaO and FeAs triple layers due to its
weakness may relax somewhat more, 0.03 \AA\ in one case.) For the La
terminated surface the relaxation is much more dramatic, it amounts to
an about 0.3 \AA\ movement of the O position outwards from the center of
the La tetrahedra. This continues into changes of about 0.03 \AA\ of the
Fe-As layer distances and even larger changes of the other inside the
slabs.

As already mentioned in Section~I, also the layer charges were compared
with their bulk values in order to assess the quality of approaching the
situation of a bulk crystal surface. As typical results, in
Table~\ref{tab:charges} the GGA layer charges as obtained by FPLO are
displayed for a bulk crystal, a cleaving crystal as described in
subsection~A, several As terminated slabs and for
(LaO)$_{10}$(FeAs)$_8$. As one can see, the charge per unit cell of the
central triple layer of the slabs deviates less than 0.01 electron
charges from the bulk value, while for the unsymmetric slab of the
cleavage situation the deviations are larger due to the additional
electric field produced by the non-zero gross dipole density.

\begin{figure}[tbh]
  \centering
  \includegraphics[scale=0.7,clip=]{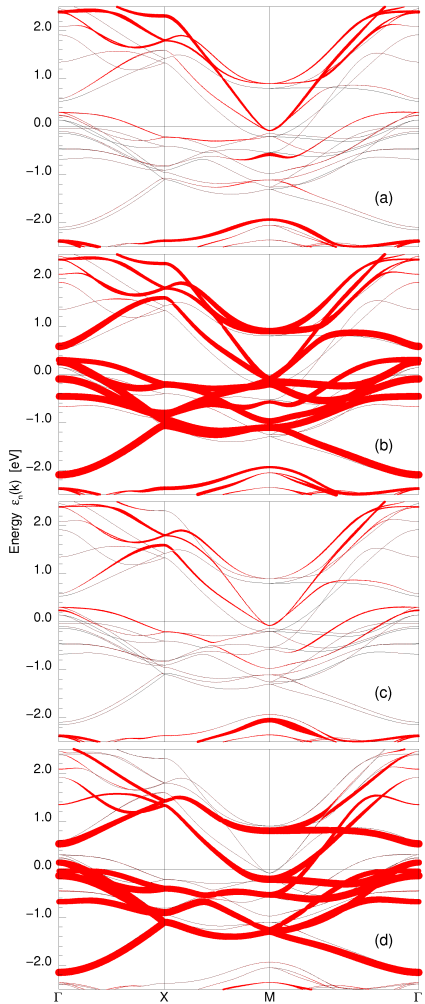}
  \caption{(color online) Band structure of the (LaO)$_4$(FeAs)$_6$
    slab. The thickness of the red (shadow) lines represents in turn the orbital
    weight of (a) surface As $4p$, (b) subsurface Fe $3d$, (c)
    subsurface As $4p$, and (d) central (bulk) Fe $3d$.}
  \label{fig:Asbands}
\end{figure}

\subsection{The As terminated surface}

As can be inferred from Tables~\ref{tab:dist} and \ref{tab:charges}, the
central Fe-As triple layer of (LaO)$_4$(FeAs)$_6$ is already close to a
bulk-like state. Also the Fermi level relative to the bands derived from
this central layer is bulk-like. Therefore, the band structure of this
slab shown in Fig.~\ref{fig:Asbands} may be considered as representative
for a bulk crystal with an As terminated (001) surface. Observe that the
slab bands do not have $k_z$-dispersion due to the slab confinement of
the Kohn-Sham states. As is seen in Fig.~\ref{fig:bulkbands}, the
$k_z$-dispersion of the bulk bands crossing Fermi level is also
negligible. Therefore, in the following only the in-plane dispersion of
the bands is discussed and presented in the figures. 

\begin{figure}[tbh]
  \centering
  \includegraphics[scale=0.6,clip=]{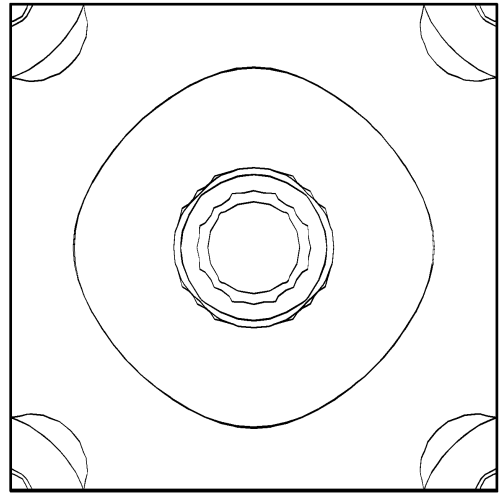}
  \caption{The 2D FSs of the slab (LaO)$_4$(FeAs)$_6$. The zone center
    is $\Gamma$ and the corner point is $M$. From $\Gamma$ outward in
    turn the FSs are: 2 times bulk, 3 times surface; from $M$ outward:
    two tiny surface FSs, two bulk FSs. (The small wiggles are due to
    the resolution of the used $\bm k$~mesh.)}
  \label{fig:FSAs}
\end{figure}

By the thickness of colored lines, the orbital character of the bands is
indicated in Fig.~\ref{fig:Asbands} from top to bottom for basis
orbitals in which the Kohn-Sham band wavefunctions are expanded: of
surface As $4p$-orbitals, Fe $3d$-orbitals of the Fe layer below the
surface As layer, As $4p$-orbitals of the next layer below this Fe
layer, and finally of the Fe $3d$-orbitals in the center of the slab
which corresponds to the first Fe layer below the surface triple layer
of a bulk crystal. The latter Fe atoms are about 10 \AA\ below the
position of the surface As atoms.

The Kohn-Sham band wavefunctions in the vicinity of Fermi level are
formed by the above accounted `chemical basis' orbitals to about 99
percent so that the thick colored lines of Fig.~\ref{fig:Asbands} (and
also of Fig.~\ref{fig:Labands} below) completely represent the extension
of the corresponding Kohn-Sham wavefunctions. Polarization states, and
other basis states besides the explicitly discussed, of the full basis
used in the calculations to not contribute to these Kohn-Sham
wavefunctions.

Both the surface Fe $3d$-bands and the bulk Fe $3d$-bands cross the
Fermi level, however, they form quite different Fermi surfaces (FS) as
shown in Fig.~\ref{fig:FSAs}. Around $\Gamma$ there are two hole
cylinders of bulk bands and around $M$ there are two electron cylinders,
all much like in a bulk crystal calculation without surface. Note that
in the slab (LaO)$_4$(FeAs)$_6$ on which Fig.~\ref{fig:FSAs} is based
the surfaces FSs are (almost) twofold degenerate due to the two surface
FeAs triple layers on both sides of the slab (which do almost not
interact due to the $z$-confinement of all conduction states). Hence,
each of the surface FSs is doubly degenerate which is, however, not
resolved in the figure.

\begin{figure}[tbh]
  \centering
  \includegraphics[scale=0.7,clip=]{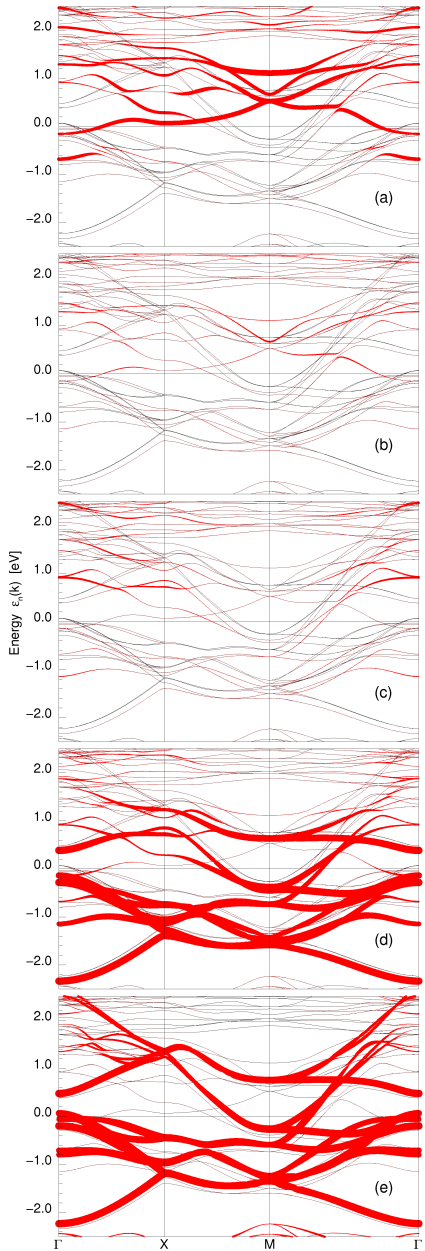}
  \caption{(color online) Band structure of the (LaO)$_{10}$(FeAs)$_8$
    slab. The thickness of the red (shadow) lines represents in turn the
    orbital weight of (a) surface La $5d+6s$, (b) subsurface O $2p$, (c)
    subsurface La $5d+6s$, (d) next to surface Fe $3d$, and (e) inner
    (bulk) Fe $3d$.}
  \label{fig:Labands}
\end{figure}

There is next to no contribution of orbitals in the subsurface LaO
triple layer as well as in any LaO triple layer to these bands (cf.\
Fig.~\ref{fig:bulkbands}). There the Kohn-Sham wavefunction amplitude is
already reduced by about an order of magnitude.

As seen in panel (b) of Fig.~\ref{fig:Asbands}, there are three surface
bands forming FS hole cylinders around $\Gamma$. Two of them have radii
only slightly larger than the bulk FSs but one has a radius about twice
as large. It results from a band of Fe $3d_{z^2}$ orbital character
which is completely occupied in the bulk. Since all conduction bands of
these materials are Fe-As antibonding,\cite{tb} it is shifted up by the
surface compression of the As tetrahedra in $z$-direction strengthening
the FeAs covalency of the $3d_{z^2}$ orbital with the As $4p$ orbitals
around $\Gamma\; (\bm k = 0)$.

There are again two electron cylinders around $M$, but they are tiny
having radii about three times smaller than the corresponding bulk FSs.
The surface states are essentially only formed by Fe $3d$-orbitals of
the surface triple layer. They have very small amplitudes everywhere
else.

\subsection{The La terminated surface}

Table~\ref{tab:dist} says that the surface relaxation of an LaO surface
triple layer is dramatic. While the geometry of the La tetrahedra
changes little, the O sites move outward by about 0.3 \AA. The charge per
unit cell of the O layer changes by 0.3 electron charges compared to the
bulk while in the case of an FeAs surface triple layer the corresponding
change of the Fe charge was only within 0.02 electron charges. The
subsurface La layer charge still changes by 0.2 electron charges which
induces even some change in the next FeAs triple layer and only beneath
that a near bulk situation is found. In contrast, only the charge of the
topmost As atom layer differs noticeably from the bulk value in the case
of the As terminated surface.

\begin{figure}[tbh]
  \centering
  \includegraphics[scale=0.6,clip=]{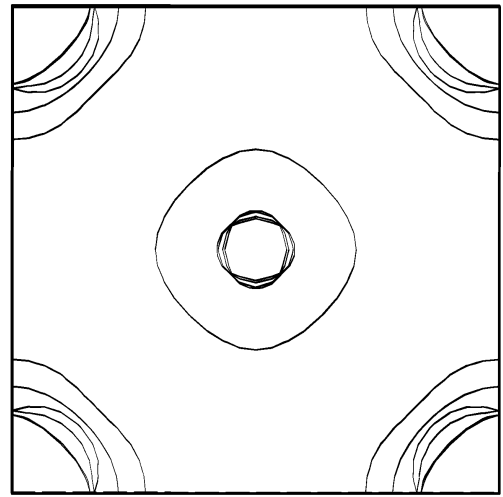}
  \caption{The 2D FSs of the slab (LaO)$_{10}$(FeAs)$_8$. Compare
    Fig.~\ref{fig:FSAs}. The FSs from $\Gamma$ outward are: two times
    bulk Fe $3d$, the outer one surface La $6s$; from $M$ outward: two
    times bulk Fe $3d$, the second one split in the slab, see text for
    explanation, two outer subsurface Fe $3d$ FSs.}
  \label{fig:FSLa}
\end{figure}

Correspondingly dramatic are the changes in the surface electronic
structure presented on Fig.~\ref{fig:Labands}. In contrast to the inner
LaO triple layers the topmost La layer is conducting: The electron
attractive potential of the O ions moving closer to the
topmost La sites brings the outermost La $6s$-bands down in
energy by about 2~eV compared to its bulk position so that it crosses
Fermi level forming a large electron cylinder FS around $\Gamma$,
see. Fig.~\ref{fig:FSLa}. This is another case of the strong
polarizability of oxygen in transition metal oxide compounds related to
the resonant instability of the O$^{2-}$ ion.  

As Fig.~\ref{fig:FSLa} is based on the results for the slab
(LaO)$_{10}$(FeAs)$_8$, all FSs shown are again twofold degenerate, and
a noticeable splitting is only seen for the second FS around $M$ due to
a small interaction of the subsurface FeAs triple layers via the next
deeper FeAs layers. Note also that the `bulk' FSs of the figure are
slightly shrunk around $\Gamma$ and expanded around $M$ compared to true
bulk FSs. This is because the slab (LaO)$_{10}$(FeAs)$_8$ is still not
thick enough for the Fermi level to converge to its bulk value. There
would, however, be no change of the character of the results any more
for even thicker slabs.

The wavefunction of the La surface band is essentially decayed already
at the neighboring O layer and there is a barrier for states at Fermi
level down to the next Fe layer (about 4~\AA\ wide). This Fe layer still
has strongly deformed bands compared to the bulk with no FS around
$\Gamma$ and two quite large hole cylinders around $M$. Only at the next
deeper Fe layer (about 14~\AA\ beneath the crystal surface) the
wavefunction of the bulk bands starts off.

Hence, the subsurface FeAs triple layer bands of an La terminated
surface deviate just in the opposite direction from the bulk compared to
the surface FeAs triple layer bands of an As terminated surface: they
shrink to zero at $\Gamma$ and expand substantially around $M$. In
addition a conducting surface La $6s$-band forms a large electron
cylinder around $\Gamma$.

\section{Summary and Conclusions}

Due to the layered structure of the recently discovered iron based
superconductor materials, their electronic structure is to various
degrees quasi-2D, most pronounced in LaOFeAs, the material considered in
this paper. As expected this situation gives rise to a well developed
and perpendicular to the crystal surface well localized surface band
structure which appears to be dramatically different from that in the
bulk.

Closely interrelated with this deviating surface electronic structure
are quite strong surface relaxations of the crystal structure, in the
considered material most dramatic for the La terminated surface.
Numerical structure relaxations reveal that LaOFeAs has quite stiff LaO
and FeAs triple layers as structural subunits which are bound together
in a much weaker way. As a result, tensile stress in $z$-direction
practically only expands the distance between these subunits and hardly
changes these substructure bond lengths themselves. This is why in a
cleavage process very coherent As and La terminated surfaces are
expected only.

Since the structural subunits are mutually charged, and those charges
are separated in the cleavage process, the strong surface relaxations
are even enhanced and the surface triple layers now also internally
relax mainly due to their change in charging. While the As tetrahedra
shrink in $z$-direction keeping the Fe ions close to their centers, the
La tetrahedra remain nearly unchanged, but the O ions move out of their
centers by about 0.3 \AA. This is related to a strong change of the
ionization state of the O ion, while the ionization state of the Fe ions
is hardly changed at the surface. Only the topmost As ions change their
ionization state strongly as expected due to the missing doping
counterpart, and, of course, also like the topmost La ions of the La
terminated surface. 

Both the As and La terminated surfaces develop a 2D surface band
structure of in $z$-direction well confined band states while the bulk
band states are dying off between 10 and 8 \AA\ below the surface As
sites of an As terminated surface and even about 4 \AA\ deeper in the
case of an La terminated surface. Both surface band structures develop
FSs. For the As terminated surface and compared to bulk these FSs are
expanded around $\Gamma$ and shrunk nearly to zero at $M$. Not really a
surprise, an additional FS appears of a band of predominantly Fe
$3d_{z^2}$ orbital character (which is completely occupied in the
bulk). It is another hole FS with a large radius around $\Gamma$ and,
due to its orbital character its wavefunctions are closest to the
surface. For the La terminated surface, on the opposite, the FSs around
$\Gamma$ disappear completely for the surface bands while those around
$M$ are expanded. However, as a real surprise, a new electron FS around
$\Gamma$ appears with a medium radius, which has La $6s$ orbital
character. Towards zone boundary the orbital character of this band
continuously changes into mainly La $5d_{xz}$ and $5d_{yz}$. Compared to
the bulk the corresponding states are lowered in energy by about 2 eV by
forming a most strong covalent bond with the O neighbor. It is this
energy gain which moves the O ion out of the center of the La
tetrahedron.

These surface band structures must be expected strongly to influence any
surface sensitive experimental result. In particular, it seems to us
that ARPES results for NdOFeAs\cite{ARPES1} (doped with F) as well as of
LaOFeAs\cite{ARPES2} show tendencies of As terminated surfaces.

\begin{acknowledgements} 

  We gratefully acknowledge assistance by U. Nitzsche with the use of
  computer facilities. We thank A. Koitzsch and J. van den Brink for
  helpful discussions.

\end{acknowledgements}

\end{document}